\documentclass{article}
\usepackage{spconf,amsmath,graphicx,hyperref}

\usepackage{tikz}
\usetikzlibrary{shapes,arrows}

\newcommand{\round}[1]{\ensuremath{\lfloor#1\rceil}}

\title{PSEUDO-CEPSTRUM: PITCH MODIFICATION FOR MEL-BASED NEURAL VOCODERS}
\name{
\begin{tabular}{c}
	Nikolaos Ellinas$^1$,
	Alexandra Vioni$^1$,
	Panos Kakoulidis$^1$,
	Georgios Vamvoukakis$^1$,
	Myrsini Christidou$^1$,\\
	Konstantinos Markopoulos$^1$,
	Junkwang Oh$^2$,
	Gunu Jho$^2$,
	Inchul Hwang$^2$,\\
	Aimilios Chalamandaris$^1$,
	Pirros~Tsiakoulis$^1$
\end{tabular}
}
\address{$^1$Innoetics, Samsung Electronics, Greece\\
	$^2$Mobile eXperience Business, Samsung Electronics, Republic of Korea}

\begin{document}
\ninept
\maketitle
\begin{abstract}
This paper introduces a cepstrum-based pitch modification method that can be applied to any mel-spectrogram representation.
As a result, this method is compatible with any mel-based vocoder without requiring any additional training or changes to the model.
This is achieved by directly modifying the cepstrum feature space in order to shift the harmonic structure to the desired target.
The spectrogram magnitude is computed via the pseudo-inverse mel transform, then converted to the cepstrum by applying DCT.
In this domain, the cepstral peak is shifted without having to estimate its position and the modified mel is recomputed by applying IDCT and mel-filterbank.
These pitch-shifted mel-spectrogram features can be converted to speech with any compatible vocoder.
The proposed method is validated experimentally with objective and subjective metrics on various state-of-the-art neural vocoders as well as in comparison with traditional pitch modification methods.
\end{abstract}
\begin{keywords}
pitch shifting, neural pitch modification, cepstrum, mel-spectrum, pseudo-cepstrum
\end{keywords}
\section{Introduction}
\label{sec:intro}

Speech synthesis applications, such as text-to-speech (TTS) and voice conversion (VC),
have been revolutionized by the advent of deep learning,
leading to significant improvements in the naturalness and intelligibility of generated audio. 
Modern approaches predominantly employ a two-stage pipeline.
First, the generation and processing is performed in an intermediate acoustic representation.
Then, a vocoder synthesizes the audio waveform from this representation.
Among various possible intermediate representations, the mel-spectrogram has emerged as the de facto standard.

Early methods for mel-spectrogram inversion, such as the Griffin-Lim algorithm \cite{griffinlim1984}, often introduced audible artifacts.
Neural vocoders have greatly overcome these limitations, and can generate high-fidelity speech.  
These models, include autoregressive architectures like WaveNet \cite{wavenet} and WaveRNN \cite{wavernn},
flow-based models like WaveGlow \cite{waveglow}, 
Generative Adversarial Networks (GANs) like HiFiGAN \cite{hifigan} and BigVGAN \cite{bigvgan},
and diffusion-based models like DiffWave \cite{diffwave} and WaveGrad \cite{wavegrad}.

Controllability of the generated audio is also an important aspect of speech synthesis applications, including volume, speaking rate and pitch control.
Traditional methods based on digital signal processing (DSP)
can perform rate and pitch manipulation directly on the speech signal
or via an analysis and re-synthesis manner.
These include the time-domain pitch synchronous overlap add (TD-PSOLA) algorithm \cite{psola},
the Harmonic plus Noise Model (HNM) \cite{hnm},
and the STRAIGHT \cite{straight} and WORLD \cite{world} vocoders.
TD-PSOLA is still considered state-of-the-art in pitch manipulation 
when the modification factor is within the operating range \cite{morrison2020controllable}.

While some speech synthesis models incorporate implicit or explicit controllability modules \cite{fastpitch, vits, ellinas2023controllable},
many rely on the vocoder for this task.
Indeed, controllable neural vocoders is active research field,
and many techniques have been proposed. 
The LPCNet vocoder appends pitch and correlation features to the bark-scale cepstral coefficients \cite{lpcnet},
and was further adapted for pitch-shifting and time-stretching in \cite{morrison2021neural}. Neural source-filter based models directly incorporate
$F0$ contours as an input conditioning signal \cite{wang2019neural, yoneyama2023high, ohtani2025fast}.
A hider-finder-combiner GAN based architecture was proposed
in \cite{webber2020hider} as neural analysis-synthesis method
that can control various speech parameters including pitch.
The pitch-shifting WaveNet vocoder proposed in \cite{morrison2020controllable}
is conditioned on the $F0$ contour and 21-channel mel-cepstral coefficients.
Pitch-dependent dilated convolution networks (PDCNNs) were proposed
that dynamically change the network architecture according to the
auxiliary $F0$ feature used as conditioning \cite{wu2021quasi}.
In a similar fashion, \cite{matsubara2023harmonic} proposes a downsampling
network for $F0$ dependent excitation signals,
and layerwise pitch-dependent dilated convolutional networks.
A HiFiGAN architecture is used in \cite{morrison2024fine},
that is conditioned on high level features such as $F0$, periodicity, loudness,
and phonetic posteriograms instead of mel-spectrogram features.
A denoising diffusion probabilistic model (DDPM) that is explicitly 
conditioned on $F0$ is proposed in \cite{hono2024periodgrad}.
This work also compares the conditioning with 50-dimensional mel-cepstral
coefficients versus 80-dimensional log mel-spectrogram features.
Their findings indicate that when the latter are used the model
utilizes pitch information embedded in the mel-spectrogram
and fails to utilize the explicit $F0$ feature for pitch shifting.
A GAN based architecture conditioned on the spectral envelope along with
the pitch contour is proposed in \cite{zhuang2021karatuner}
for the pitch correction task. Cycle-consistency GAN training was proposed in
\cite{gu2025neurodyne} following the hider-finder-combiner paradigm.
Also in the context of pitch correction task,
\cite{hai2023diff} propose a diffusion model that generates
mel-spectrum from LPC or WORLD vocoder features.



\tikzset{
	block/.style = {draw, fill=white, rectangle, minimum height=3em, minimum width=4em, text width=6em, align=center},
	tmp/.style  = {coordinate}, 
	sum/.style= {draw, fill=white, circle, node distance=1cm},
	input/.style = {coordinate},
	output/.style= {coordinate},
	pinstyle/.style = {pin edge={to-,thin,black}
	}
}

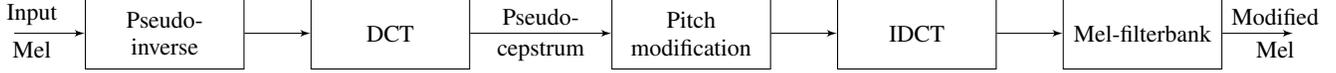
\begin{figure*}[!htb]
	\centering
	\begin{tikzpicture}[auto, node distance=2cm,>=latex']
		\node [input, name=mel] (melinput) {};
		\node [block, right of=melinput] (pinverse) {Pseudo-inverse};
		\node [block, right of=pinverse, node distance=3cm] (dct) {DCT};
		\node [block, right of=dct, node distance=4cm] (pitchmod) {Pitch modification};
		\node [block, right of=pitchmod, node distance=3cm] (idct) {IDCT};
		\node [block, right of=idct, node distance=3cm] (mel) {Mel-filterbank};
		\node [output, right of=mel] (meloutput) {};
		\draw [->] (melinput) -- (pinverse) node[near start, above]{Input} node[near start, below]{Mel};
		\draw [->] (pinverse) -- (dct);
		\draw [->] (dct) -- (pitchmod) node[midway, above]{Pseudo-} node[midway, below]{cepstrum};
		\draw [->] (pitchmod) -- (idct);
		\draw [->] (idct) -- (mel);
		\draw [->] (mel) -- (meloutput) node[near end, above]{Modified} node[near end, below]{Mel};
	\end{tikzpicture}
	\caption{The proposed pitch modification method on the mel-spectrogram domain.
		The pseudo-inverse transform followed by DCT,
		as well as IDCT followed by the mel-filterbank,
		can be combined in a single linear transformation.} \label{fig:method}
\end{figure*}

The majority of proposed neural vocoders that support pitch modification,
rely on explicit $F0$ conditioning. 
$F0$ estimation is necessary either as a part of the model, or as an external module.
Usually, smooth spectral features are preferred over spectral representations that contain $F0$ information.
In this work, we propose the first DSP based approach that applies 
pitch shifting directly on the mel-spectrogram domain.
Thus, unlocking pitch shifting capabilities for any neural vocoder
that is based on mel-spectrogram features, without the need for
any additional training or modifications in the model architecture.
The method utilizes mel pseudo-inversion followed by DCT
to convert the features to the cepstral domain.
Pitch is then directly shifted in the cepstral domain,
and the features are converted back to the mel-spectral domain (Fig.~\ref{fig:method}).
The cepstral peak position corresponding to pitch is not required,
nor $F0$ estimation is necessary,
avoiding any degradation introduced by $F0$ or voicing errors.
Moreover, the whole process is applied at the frame level
enabling fine-grained control, while being extremely lightweight
and essentially not affecting the runtime factor or latency of the vocoder.

\section{Proposed method}
\label{sec:method}

\subsection{Cepstrum}
\label{ssec:cepstrum}

\textit{Cepstrum} is defined as the inverse Fourier transform
of the log magnitude spectrum of a signal,
and is a special case homomorphic filtering
\cite{rabiner2007introduction}.
The cepstrum of a discrete-time signal is defined as
\begin{equation}\label{eq:cepstrum}
	c[n] = \frac{1}{2\pi} \int_{-\pi}^{\pi} \log|X(e^{j\omega})|e^{j\omega n} d\omega 
\end{equation}
where $X(e^{j\omega})$ is the discrete-time Fourier Transform (DTFT) of the signal $x[n]$.
The \textit{complex cepstrum} is defined by replacing the logarithm of the magnitude
with the complex logarithm of $X(e^{j\omega})$
\begin{equation}\label{eq:complex_cepstrum}
	\hat{x}[n] = \frac{1}{2\pi} \int_{-\pi}^{\pi} \log{X(e^{j\omega})}e^{j\omega n} d\omega 
\end{equation}

In practical applications, such as digital signal processing of speech signals,
the cepstrum is computed on a frame basis by replacing the DTFT with the
Discrete Fourier Transform (DFT) af follows
\begin{equation}\label{eq:dft}
	X[k]=\sum_{n=0}^{N-1}x[n]e^{-j(2\pi k/N)n}
\end{equation}
\begin{equation}\label{eq:complex_dft}
	\hat{X}[k]=\log|X[k]| + j\arg{X[k]}
\end{equation}
\begin{equation}\label{eq:complex_discrete_cepstrum}
	\tilde{\hat{x}}[k]=\frac{1}{N}\sum_{n=0}^{N-1}\hat{X}[k]e^{j(2\pi k/N)n}
\end{equation}
\begin{equation}\label{eq:discrete_cepstrum}
\tilde{c}[k]=\frac{1}{N}\sum_{n=0}^{N-1}\log|X[k]|e^{j(2\pi k/N)n}
\end{equation}
where $N$ is the length of sampled signal $x[n]$,
$X[k]$ is its DFT, $\hat{X}[k]$ is the complex logarithm $X[k]$,
$\tilde{c}[k]$ is the cepstrum,
and $\tilde{\hat{x}}[k]$ is the complex cepstrum of $x[n]$.

The main advantage of the cepstrum representation is the clear separation
of the source and filter components of voiced speech.
The indexing term $k$ of the cepstrum, also referred as quefrency,
represents a measure of time that measures the periodicity of frequencies
in the spectral domain. 
The low quefrency part of the cepstrum corresponds to the slow variations 
(with respect to frequency) in the log spectrum,
while the high quefrency components correspond to the more rapid fluctuations
of the log spectrum.
This characteristic of the cepstrum domain, enabled its application
for various tasks, such as format and pitch tracking.
This is also the main inspiration for this work, since pitch is represented
by a clear cepstral peak, which we propose to directly shift it in the quefrency domain.
However, this is not directly applicable, since even in the case of the complex cepstrum
that there exists an inverse transform,
a simple shift would introduce phase discontinuities.
Such attempts have been proposed in the past that require
pitch-synchronous cepstral analysis \cite{maia2013complex, seiyama1992new}.

\subsection{Mel Pseudo-cepstrum}
\label{ssec:melcep}

The cepstral analysis presented above is applied directly on the log magnitude domain.
However, for the task at hand, we need to adapt it to the mel log-spectrogram domain.
The mel log-spectrum is calculated via a linear transformation of the log magnitude,
i.e. via matrix notation as follows
\begin{equation}\label{eq:mel}
	\mathbf{S}=\log|\mathbf{M} \cdot \mathbf{X}|
\end{equation}
where $\mathbf{M}$ is the mel-filterbank matrix, and $\mathbf{S}$ the mel-spectrogram
features, i.e. the domain of many state-of-the-art vocoders.

The mel-frequency cepstral coefficients (MFCCs) are derived by transforming $\mathbf{S}$
via the discrete cosine transform (DCT).
However, this deviates from original cepstrum
formulation because $\mathbf{S}$ is indexed in the a log-frequency domain. 
Hence, the harmonics of $F0$ are not linearly scaled and do not produce a prominent
cepstral peak.

\begin{figure*}[htb]
	\centering
	\centerline{\includegraphics[width=\linewidth]{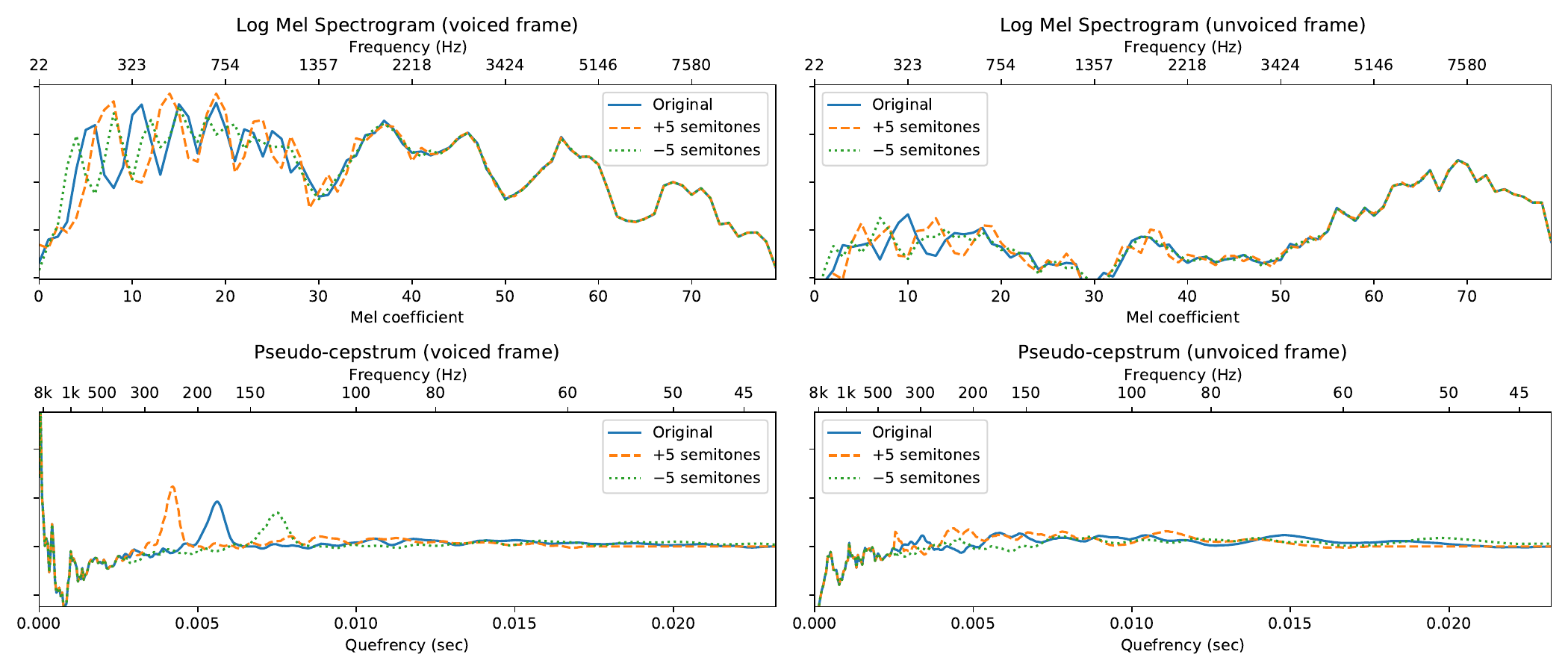}}
	\caption{Log mel-spectrogram (top) and the corresponding pseudo-cepstrum (bottom)
		for a voiced speech frame (left) as well as an unvoiced speech frame (right).
		The pitch shifted versions for $+5$ and $-5$ semitones are also shown.
		}
	\label{fig:cep}
\end{figure*}

Alternatively to the MFCC formulation, we propose to first apply the 
pseudo-inverse of the mel-filterbank transform similarly to \cite{lv2024freev}.
However, we do not de-normalize the log-compression of $\mathbf{S}$.
Instead, we apply DCT in the linear frequency domain as follows
\begin{equation}\label{eq:mfcc}
	\mathbf{C}=\mathbf{D} \cdot \mathbf{M^{+}} \cdot \mathbf{S}
\end{equation}
where $\mathbf{M^{+}}$ is the pseudo-inverse of $\mathbf{M}$,
and $\mathbf{D}$ is the DCT transformation matrix.
Our formulation also differs from the standard definition of cepstrum,
due to the inclusion of the mel-filterbank and the use of DCT instead of DFT.
However, the separation of source and filter components is preserved
, as shown in Fig.~\ref{fig:cep}.
To distinguish $\mathbf{C}$ from traditional cepstrum we define it as 
\textit{pseudo-cepstrum}.

\subsection{Pitch modification}
\label{ssec:pitchmod}

The pitch modification in the pseudo-cepstrum domain is accomplished via 
shifting of the cepstral peak. In order to achieve this, we propose 
to shift the entire source component of the pseudo-cepstrum, 
i.e. the part of the pseudo-cepstrum above a threshold quefrency
that corresponds to the maximum $F0$ value $F0_{max}$.
This is the only parameter that needs to be specified for the method.
However, the shifting in the pseudo-cepstrum domain is not done linearly
because the quefrency axis is not linear with respect to frequency.
In order to achieve a linear shift in frequency,
the pseudo-cepstrum is interpolated in the quefrency domain as described below.

The quefrency and frequency values corresponding to the $k$-th coefficient 
of a pseudo-cepstrum frame $c[k]$ are respectively given by
$q[k] = k/f_s$ and $f[k] = 1/q[k]$,
where $f_s$ is the sampling frequency in Hz.
We first define a modification vector similarly to a liftering window as follows
\begin{align}\label{eq:cep_mod_factor}
	w[k] = 
	\begin{cases}
		1 & k \leq q_{min} \\
		2^\frac{x}{12} & k > q_{min}
	\end{cases}
\end{align}
where $x$ is the pitch shift is semitones, and $q_{min}=1/F0_{max}$ is the quefrency corresponding to the maximum $F0$ threshold.
Smoothing of $w[k]$ can be applied in order to reduce the discontinuity at $q_{min}$.
The modified pseudo-cepstrum is then derived via interpolation,
e.g nearest-neighbor, linear, etc.
For nearest-neighbor interpolation the modified cepstrum frame is given by
\begin{align}\label{eq:cep_mod}
	c^\prime[k] = w[k] c\left[\round{w[k] k}\right]
\end{align}
where $\round{\cdot}$ denotes the rounding operator. 
The normalization with $w$ is applied to compensate for the stretching/shrinking
caused by the non-linear shifting of the cepstrum.

The modified pseudo-cepstrum is then converted back to the spectral domain
by applying the inverse discrete cosine transform (IDCT).
Next, the mel-filerbank transform is applied in order to get 
the modified mel-spectrogram.
The whole pipeline is shown in Fig.~\ref{fig:method}.
The method is also applied to the unvoiced frames
without significantly affecting them, since the lower part of the spectrum 
that is affected contains little audible content for unvoiced sounds.

\section{Experiments}
\label{sec:experiments}

In order to assess the effectiveness and quality of the proposed method, we conduct objective and subjective evaluation experiments by applying it across various state-of-the-art vocoders and pitch modification ranges.
For the subjective evaluation we perform crowd-sourced Mean Opinion Score (MOS) tests based on utterance naturalness.
For the objective evaluation we calculate three commonly used metrics to assess pitch shifting performance.

We use a test set of 10 randomly selected utterances from the open-source LJ speech dataset \cite{ljspeech17}, which is a common benchmark for TTS models.
All utterances are resampled to 24 kHz, to match the pre-trained vocoders' sampling rate.
Pseudo-cepstrum pitch modification is first applied at the mel-spectrogram domain and then the final audio is produced with each vocoder.
We apply a pitch shift in the range of $-12$ up to $+12$ semitones, resulting in 24 pitch shifted variants for each utterance.
We used the following pretrained neural vocoders with open-source implementations: HiFiGAN\footnote{\url{https://github.com/jik876/hifi-gan}} \cite{hifigan}, BigVGAN\footnote{\url{https://github.com/NVIDIA/BigVGAN}} \cite{bigvgan}, Vocos\footnote{\url{https://github.com/gemelo-ai/vocos}} \cite{siuzdak2024vocos} and WaveFM\footnote{\url{https://github.com/luotianze666/wavefm}} \cite{wavefm}.
The Griffin-Lim vocoder \cite{griffinlim1984} and the TD-PSOLA pitch modification method \cite{psola} are also included for comparisons.
The open-source vocoder implementations use different packages and definitions for mel extraction, but this does not affect our method, as it is universal.

\subsection{Objective evaluation}
\label{ssec:objective_evaluation}

For the objective evaluation we extract and compare $F0$ contours for each modified utterance.
As reference targets for the objective metrics, the ground truth $F0$ values are multiplied by the corresponding pitch shifting factor $2^{s/12}$, where $s$ is the pitch shift in semitones.
That way, we can have estimates for each metric, assuming a perfect ground truth pitch shifting.

\begin{figure*}[htb]
	\begin{minipage}[b]{0.33\linewidth}
		\centering
		\centerline{\includegraphics[width=\linewidth]{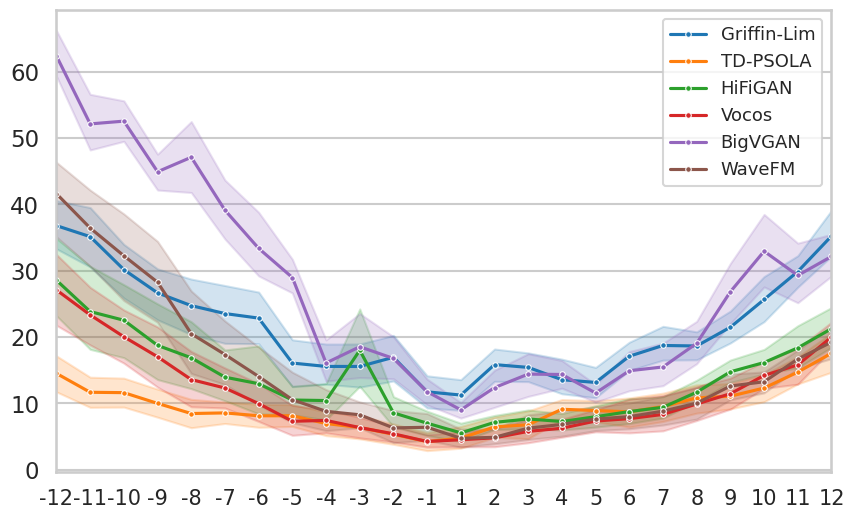}}
		\centerline{(a) FFE}\medskip
	\end{minipage}
	\begin{minipage}[b]{.33\linewidth}
		\centering
		\centerline{\includegraphics[width=\linewidth]{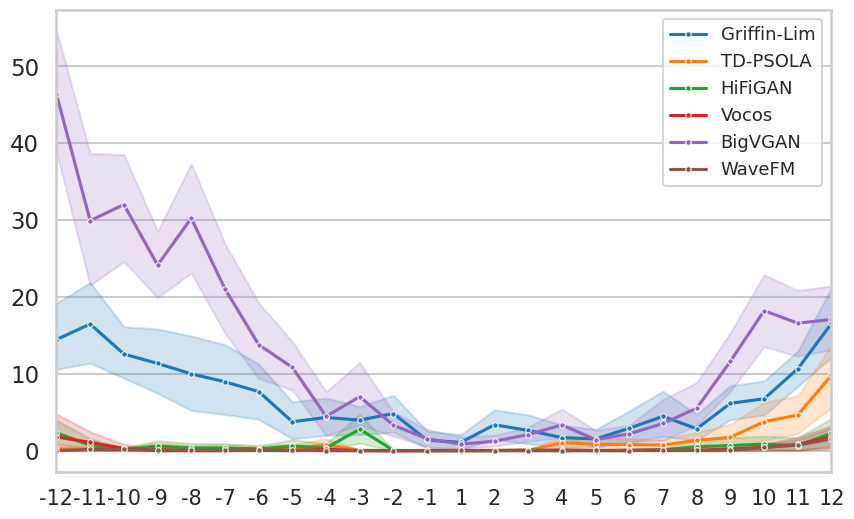}}
		\centerline{(b) GPE}\medskip
	\end{minipage}
	\hfill
	\begin{minipage}[b]{0.33\linewidth}
		\centering
		\centerline{\includegraphics[width=\linewidth]{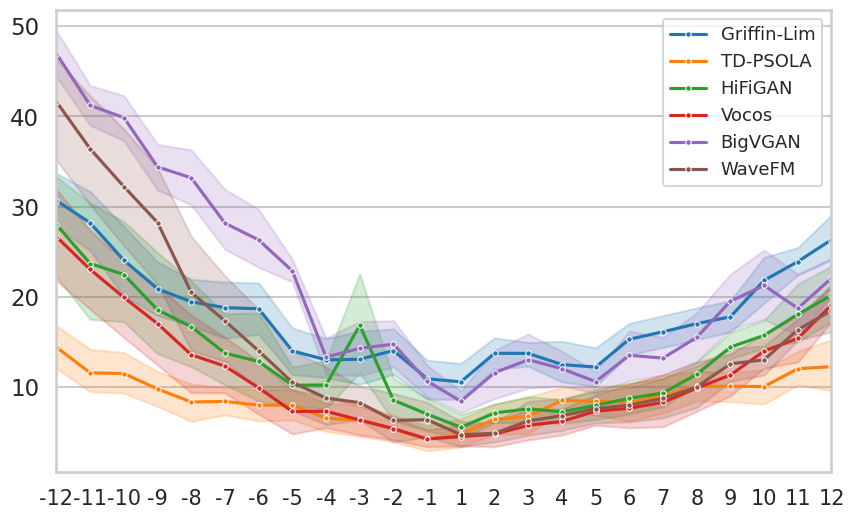}}
		\centerline{(c) VDE}\medskip
	\end{minipage}
	\vspace*{-15pt}
	\caption{Objective evaluation metrics. The horizontal axis represents the semitone shift value, whereas the vertical axis shows the value of the corresponding metric. The shadowed regions indicate the $95\%$ confidence intervals. Lower values indicate better performance.}
	\label{fig:metrics}
\end{figure*}

We use the following metrics: Gross Pitch Error (GPE), Voicing Decision Error (VDE) and $F0$ Frame Error (FFE) \cite{chu2009reducing}.
GPE measures relative errors of $F0$ higher than $20\%$ in voiced frames.
VDE is the proportion of frames where a voicing decision is wrong.
FFE is the proportion of frames where an error is made according to the 2 previous metrics and is considered the overall pitch tracking performance metric.
Temporal alignment via Dynamic Time Warping (DTW) is usually required to use these metrics, but not in this scenario since the method preserves the temporal structure of the audio.
The results for each metric are depicted in Figure~\ref{fig:metrics}.

In general, TD-PSOLA is considered a state-of-the-art method in pitch modification and this is confirmed by the very low errors especially up to moderate pitch shift ranges.
Vocos, HiFiGAN and WaveFM show the most stable results across the full modification range.
BigVGAN vocoder appears to be the most sensitive in modifying its input representations, as its performance degrades faster, especially in the negative shifts.
Finally, Griffin-Lim presents lower scores compared to most methods across more points, as by default it produces lower quality audio.

We should note that the algorithms for $F0$ estimation are also sensitive to the input and many times miscalculate the pitch values or the voiced/unvoiced regions.
Thus, the acoustic results might be better than the ones depicted in the objective metrics, especially in the tails of the modification range.
A safe and more commonly used range for TTS would be the $[-6,+6]$ semitones, where all methods have mostly stable performance.

\vspace{-3pt}

\subsection{Subjective evaluation}
\label{ssec:subjective_evaluation}

\begin{figure}[htb]
	\centering
	\centerline{\includegraphics[width=\linewidth]{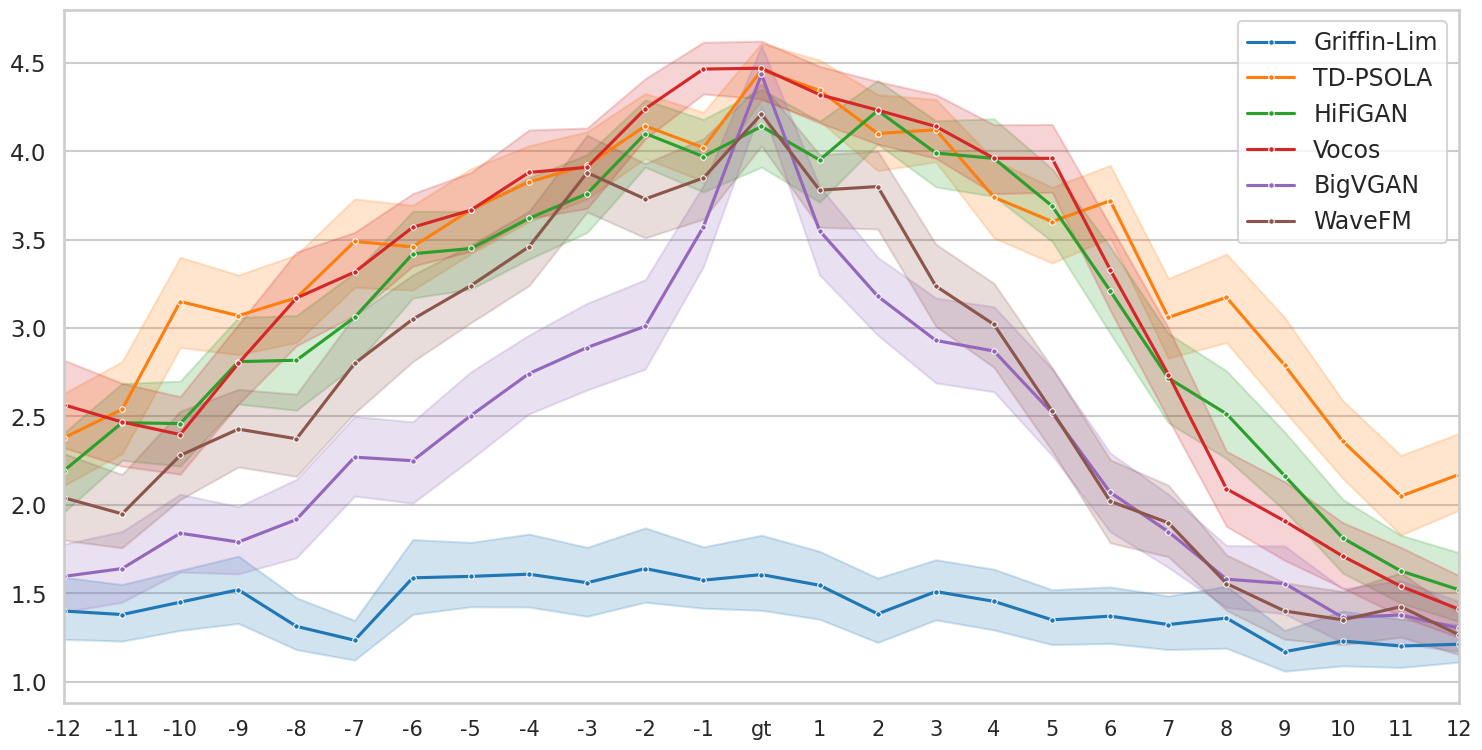}}
	\caption{Subjective listening test results. Horizontal and vertical axes represent semitone shift value and MOS score respectively. The shadowed regions indicate the $95\%$ confidence intervals.}
	\label{fig:mos}
\end{figure}

For the subjective evaluation, we performed an MOS test in order to assess naturalness for all systems across the full pitch modification range.
The test was conducted using the Prolific platform and every audio sample was evaluated by 10 unique participants.
Listeners were asked to score each utterance's naturalness on a 5-point Likert scale.
Results of the test are depicted in Figure~\ref{fig:mos}.

As expected, naturalness is the highest near the unmodified condition, where all neural vocoders score above 4.0.
By modifying the pitch in either direction, a gradual decline in MOS is observed.
The degradation is pronounced more as we approach the tails of the range, reflecting the drop in the perception of naturalness under extreme pitch alterations.

In the resulting curves we can also observe that TD-PSOLA along with HiFiGAN and Vocos score the highest, followed by WaveFM.
Since TD-PSOLA is essentially the ground truth signal modified with the overlap-add method, we can consider this to be the upper bound of all methods.
The fact that its scores overlap with two neural vocoders can be attributed to the margin of error that is inherent to all crowd-sourced tests.
From this result, we can also conclude that our proposed pseudo-cepstrum method, when combined with high quality vocoders such as HiFiGAN and Vocos, is on par with TD-PSOLA and essentially achieves a similar result in pitch modification.
BigVGAN undeperforms the rest of the neural models, revealing a reduced flexibility of the model to altered inputs.
Griffin-Lim yields the lowest scores even at ground truth utterances, which aligns with the inherent quality degradation that this method introduces.
We encourage readers to also listen to the samples at our website: \url{https://innoetics.github.io/publications/pseudocep/index.html}

\vspace*{-2pt}

\section{CONCLUSIONS}
\label{sec:conclusions}

In this work we have proposed a pitch modification method for mel-spectrograms that operates on the cepstrum domain, which is accessible via mel pseudo-inversion.
Neural vocoders achieve high-fidelity audio reconstruction from the mel-frequency domain and we believe that this addition unlocks a useful additional property.
Objective and subjective evaluations prove that our method is effective and accurate at pitch modification, as well as robust to model variation.
As future work, an alternative use of this method would be to split the cepstrum at the selected cutoff point and obtain two different representations that could be fed as two different inputs to a neural representation model and obtain disentangled representations for the vocal tract and the pitch contour.

%
%


\bibliographystyle{IEEEbib}
\bibliography{refs}

\end{document}